\documentclass[twocolumn]{aastex62}

\graphicspath{{../p_mwd/},{./figs/}}
\usepackage{comment}
\usepackage{amsmath,amssymb}
\usepackage{bm}

\newcommand{\kepler}{\textit{Kepler}}

\newcommand{\gaia}{\textit{Gaia}}

\newcommand{\msun}{M_\odot}
\newcommand{\rsun}{R_\odot}

\newcommand{\au}{\mathrm{au}}

\newcommand{\days}{\mathrm{days}}

\shorttitle{``Impossible" ELM WD in a self-lensing binary}
\shortauthors{Masuda et al.}

\begin{document}


\title{
Self-lensing Discovery of a $0.2\,M_\odot$ White Dwarf in an Unusually Wide Orbit Around a Sun-like Star\footnote{Based in part on data collected at Subaru Telescope, which is operated by the National Astronomical Observatory of Japan.}
}

\correspondingauthor{Kento Masuda}
\email{kmasuda@astro.princeton.edu}

\author{Kento Masuda}
\altaffiliation{NASA Sagan Fellow}
\affil{Department of Astrophysical Sciences, Princeton University,
Princeton, NJ 08544, USA}

\author{Hajime Kawahara}
\affil{Department of Earth and Planetary Science, The University of Tokyo, Tokyo 113-0033, Japan}
\affil{Research Center for the Early Universe, School of Science, The University of Tokyo, Tokyo 113-0033, Japan}

\author{David W. Latham}
\affil{Center for Astrophysics $|$\,Harvard \& Smithsonian, Cambridge, MA 02138, USA}

\author{Allyson Bieryla}
\affil{Center for Astrophysics $|$\,Harvard \& Smithsonian, Cambridge, MA 02138, USA}

\author{Masanobu Kunitomo}
\affil{Department of Earth and Planetary Science, The University of Tokyo, Tokyo 113-0033, Japan}

\author{Morgan MacLeod}
\altaffiliation{NASA Einstein Fellow}
\affil{Center for Astrophysics $|$\,Harvard \& Smithsonian, Cambridge, MA 02138, USA}

\author{Wako Aoki}
\affil{National Astronomical Observatory of Japan, 2-21-1 Osawa, Mitaka, Tokyo 181-8588, Japan}



\begin{abstract}

We report the discovery of the fifth self-lensing binary in which a low-mass white dwarf (WD) gravitationally magnifies its 15th magnitude G-star companion, KIC 8145411, during eclipses. The system was identified from a pair of such self-lensing events in the \textit{Kepler} photometry, and was followed up with the Tillinghast Reflector Echelle Spectrograph (TRES) on the 1.5\,m telescope at the Fred Lawrence Whipple Observatory and the High-Dispersion Spectrograph (HDS) on the Subaru 8.2\,m telescope. A joint analysis of the TRES radial velocities, the HDS spectrum, and the \textit{Kepler} photometry of the primary star determines the WD mass $0.20\pm0.01\,M_\odot$, orbital semi-major axis $1.28\pm0.03\,\mathrm{au}$, and orbital eccentricity $0.14\pm0.02$. 
Because such extremely low-mass WDs cannot be formed in isolation within the age of the Galaxy, their formation is believed to involve binary interactions that truncated evolution of the WD progenitor. However, the observed orbit of the KIC 8145411 system is at least ten times wider than required for this scenario to work. The presence of this system in the \kepler\ sample, along with its similarities to field blue straggler binaries presumably containing WDs, may suggest that some 10\% of post-AGB binaries with Sun-like primaries contain such anomalous WDs.

\end{abstract}




\section{Introduction}


The lowest-mass known white dwarfs (WDs) in the Galaxy have masses
below $0.2\,\msun$ \citep[e.g.,][]{2007ApJ...660.1451K}. Because the Galaxy is not old enough to produce them through single-star evolution, these so-called extremely low-mass (ELM) WDs with $\lesssim0.25\,\msun$ are considered to be remnants of mass transfer during the red-giant branch (RGB) phase, in which hydrogen envelope of the WD progenitor was stripped to leave a degenerate helium core with $\lesssim0.45\,\msun$, before the star ignites helium and enters the asymptotic giant branch (AGB) phase \citep[e.g.,][]{1995MNRAS.275..828M}.

The observations so far appear to be consistent with this scenario. For low-mass WDs identified as companions to millisecond pulsars, the mass transfer model successfully explained the correlation between their orbital periods and WD masses \citep{1995MNRAS.273..731R, 1999A&A...350..928T}, as well as the rejuvenated nature of the pulsars. Low-mass WDs have also been identified from the Sloan Digital Sky Survey (SDSS) spectroscopically \citep{2004ApJ...606L.147L}, and the dedicated radial-velocity (RV) companion search \citep[ELM survey;][]{2010ApJ...723.1072B} has found presumably degenerate companions on short-period ($\lesssim$ a few days) orbits for most of the surveyed WDs with $\lesssim0.25\,\msun$ \citep{2010ApJ...723.1072B, 2011ApJ...727....3K}. 
Low-mass WDs, or their precursor sub-dwarf stars, have also been identified as secondary companions to early-type dwarfs on $\lesssim10$-day orbits via (combination of) eclipses, relativistic beaming, and ellipsoidal variations in the photometric data both from ground and space \citep[e.g.,][]{2010ApJ...715...51V, 2013Natur.498..463M, 2015ApJ...815...26F}. The orbital periods and WD masses of these non-pulsar systems are also consistent with RGB mass transfer (cf. Section \ref{sec:impossible}).

This Letter reports the discovery of an ELM WD eclipsing the stellar companion KIC 8145411 separated by $\sim1\,\mathrm{au}$, which serves as an exception to the above cases. The WD was identified from periodic brightening of KIC 8145411 due to in-eclipse microlensing in the archival photometry of the \kepler\ spacecraft \citep{2010Sci...327..977B}.  This ``self-lensing" has been used to detect four WDs in binaries with orbital periods $88$--$728\,\days$ \citep{2014Sci...344..275K, 2018AJ....155..144K}, and here we report the fifth such case. We characterize the system combining the \kepler\ photometry and ground-based spectroscopy (Section \ref{sec:data}), and make a strong case that the companion is an ELM WD even though its light is not observed (Section \ref{sec:model}). We discuss the puzzle presented by this system (Section \ref{sec:impossible}), as well as potential evidence that ELM WDs are not rare among post-interaction binaries with au-scale orbits, including so-called field blue straggler (FBS) binaries (Section \ref{sec:fbs}). 
We also briefly comment on possible formation paths (Section \ref{sec:discussion}).

\section{Photometric and Spectroscopic Data}\label{sec:data}

\subsection{\kepler\ Light Curves}\label{ssec:data_kepler}

In \citet{2018AJ....155..144K}, we identified KIC 8145411 as a self-lensing binary candidate exhibiting two pulses separated by $\approx910\,\days$ in the \kepler\ pre-search data conditioning (PDC) light curve.  Since the data in the middle of the two pulses were missing, the orbital period was uncertain by a factor of two. We resolve this ambiguity with the RV data presented in this study.

In the following, we mainly use the PDC light curves within $\approx2\,\days$ around the detected pulses, after iterative detrending with a third-order polynomial function of time as described in \citet{2018AJ....155..144K}. We also analyze the simple aperture photometry (SAP) flux to check the robustness of the results against the adopted data set. The cadences are $29.4$~minutes in both cases.

\subsection{RVs from FLWO/TRES}\label{ssec:data_rv}

We obtained $12$ high-resolution ($R\sim44,000$) spectra of KIC 8145411 using the Tillinghast Reflector Echelle Spectrograph \citep[TRES;][]{2007RMxAC..28..129S, 2011ASPC..442..305M} on the 1.5\,m telescope at the Fred Lawrence Whipple Observatory (FLWO) in Arizona. The typical exposure time was $\sim1\,\mathrm{hr}$ and the signal-to-noise per resolution element was $\sim17$. The relative RVs were measured by cross correlating each spectrum against the spectrum with the highest signal-to-noise, as listed in Table \ref{tab:rvdata}. 

\begin{deluxetable}{ccc}[!ht]
\tablecaption{TRES radial velocities.\label{tab:rvdata}}
\tablehead{
\colhead{Time ($\mathrm{BJD_{TDB}}$)} & \colhead{Radial velocity (m/s)} & \colhead{Error (m/s)}
}
\startdata
2457900.8368 &  $-3575.3$ &  $123$ \\
2457935.7773 &  $-4867.5$ &  $103$\\
2458007.6608 &  $-4657.4$ & $87$\\
2458021.7223 &  $-3667.2$ & $85$\\
2458067.6248 &  $-244.5$   &  $67$\\
2458210.9310 &  $2914.0$  & $95$\\
2458256.9529 &  $627.5$    & $87$\\
2458272.9336 &  $0$        & $88$\\
2458292.8412 &  $-1058.1$ & $88$\\
2458390.6203 & $-5035.2$ & $81$\\
2458419.6013 & $-5641.4$ & $108$\\
2458439.6326 & $-4990.9$ & $191$\\
\enddata
\end{deluxetable}

\subsection{High-resolution Spectrum from Subaru/HDS}\label{ssec:data_hds}

To characterize the primary star, we also obtained a higher signal-to-noise spectrum of KIC 8145411 with the High-Dispersion Spectrograph \citep[HDS;][]{2002PASJ...54..855N} installed on the Subaru 8.2\,m telescope. We used the standard I2a set-up and the Image Slicer \#2 \citep[][$R\sim80,000$]{2012PASJ...64...77T}, and obtained one $30$-minute exposure. The resulting signal-to-noise was $\sim30$ per pixel. 

\section{Physical Model of the System} \label{sec:model}

\begin{figure*}
	\gridline{\fig{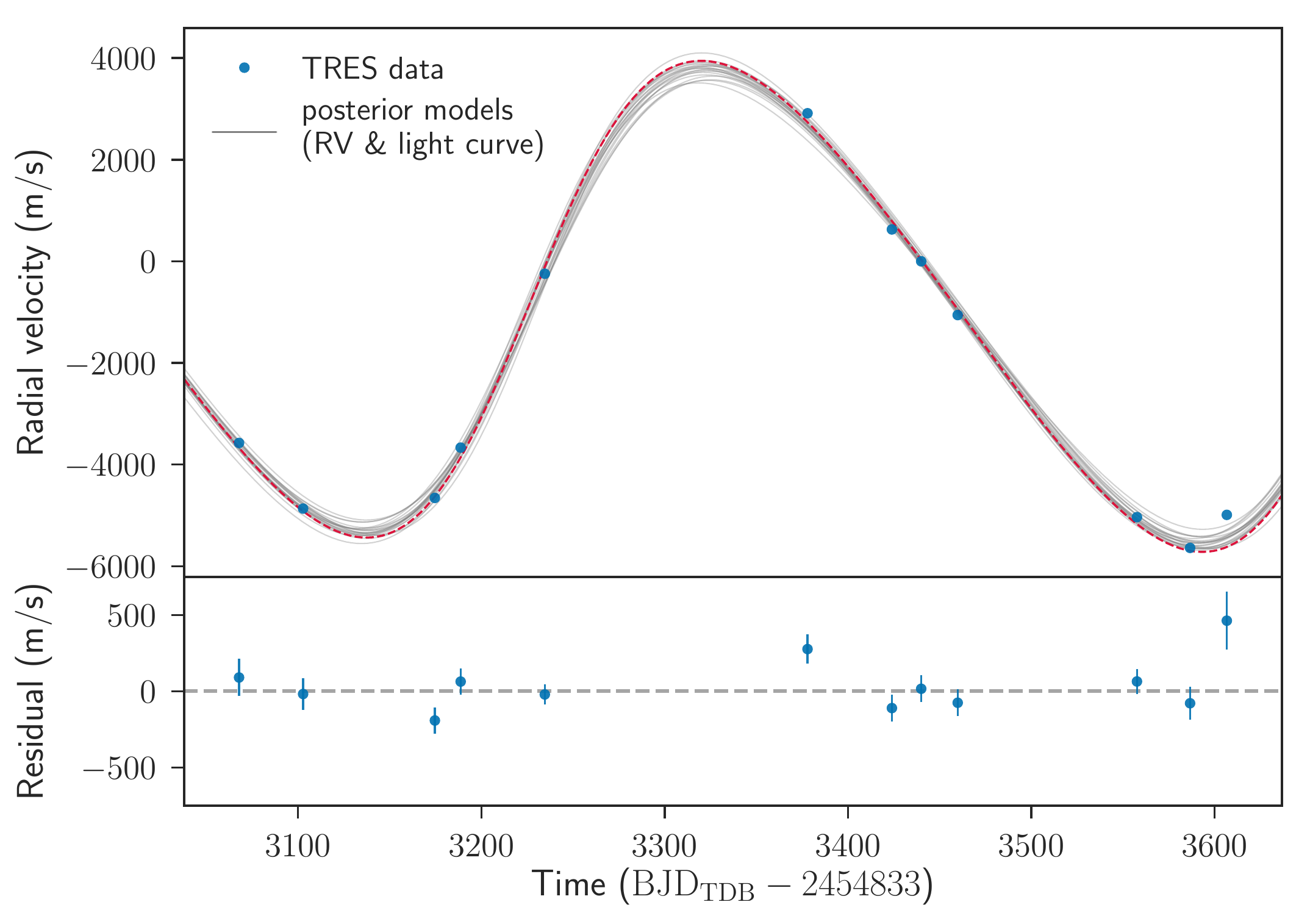}{0.7\textwidth}{(a) Radial velocity}}
	\gridline{\fig{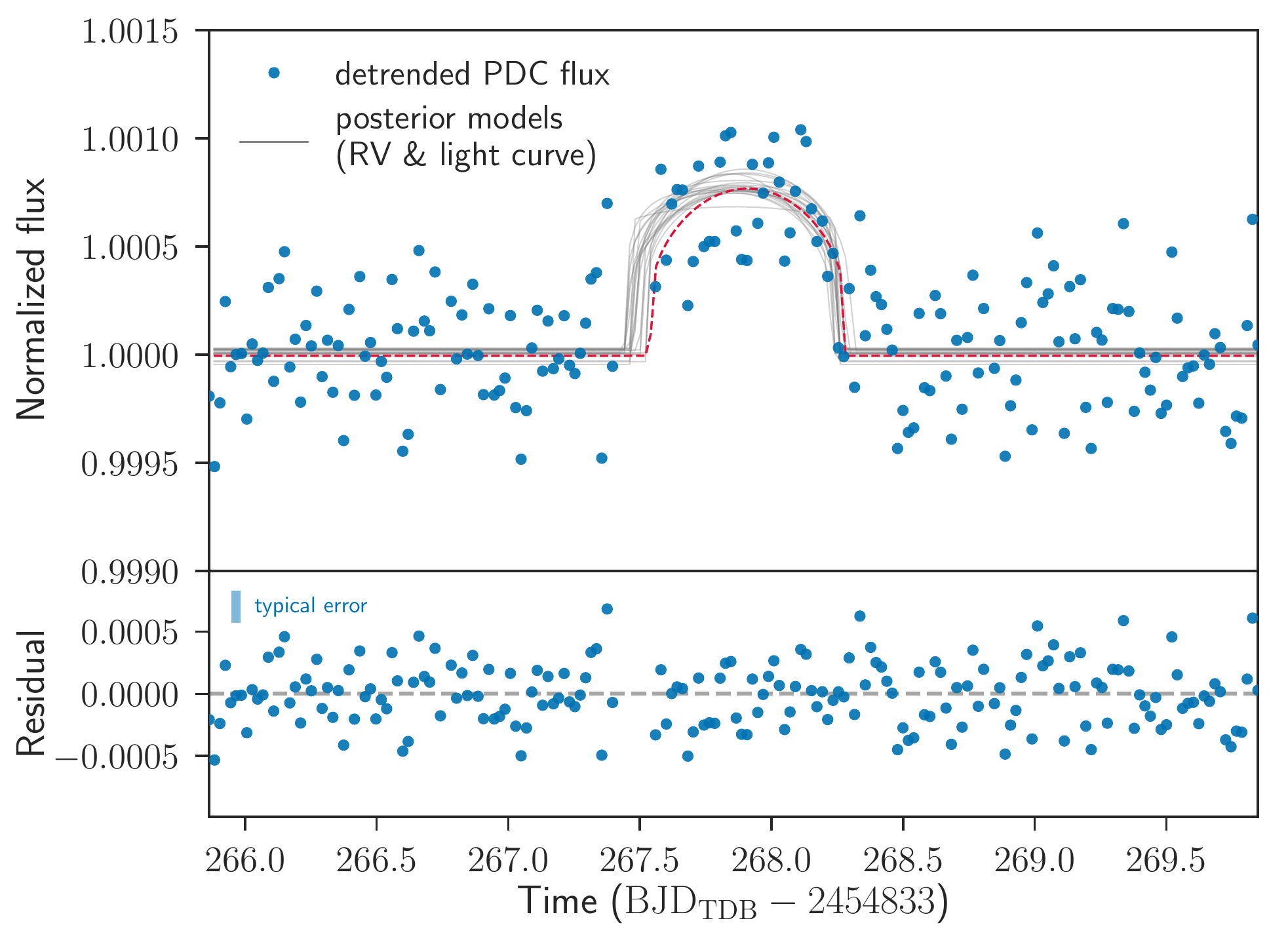}{0.505\textwidth}{(b) First observed pulse}\fig{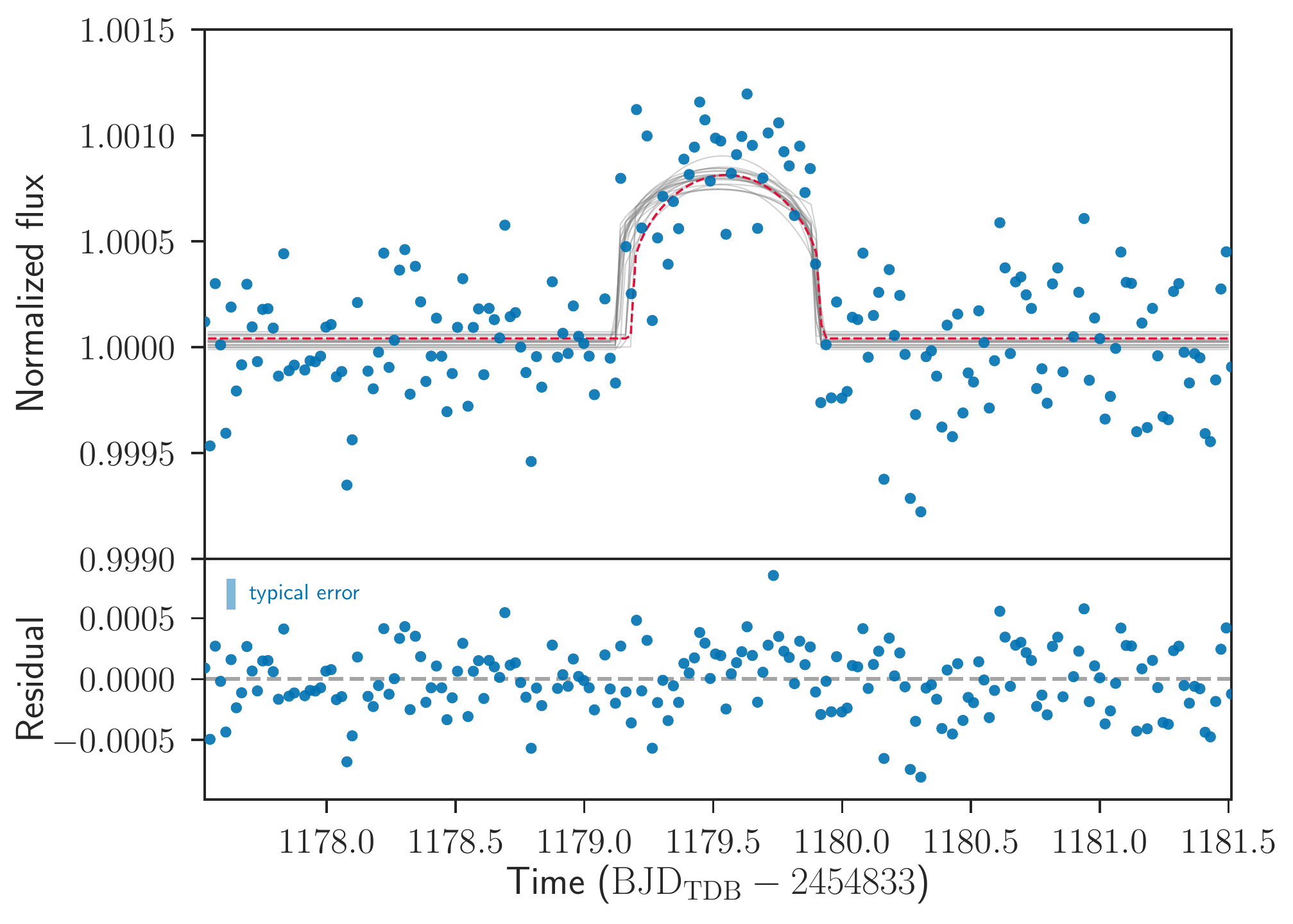}{0.505\textwidth}{(c) Second observed pulse}}
	\caption{The data (blue points) and random 20 posterior models (gray lines) from the joint analysis in Section \ref{ssec:model_joint}. The red dashed lines show the best-fit model from the analysis assuming the WD mass--radius relation (see Section \ref{ssec:model_joint}). (a) Radial velocity time series from FLWO/TRES. (b, c) detrended \kepler\ PDC light curves around self-lensing pulses.}
	\label{fig:lcrv}
\end{figure*}

The new RV data from FLWO/TRES (Section \ref{ssec:data_rv}, Figure \ref{fig:lcrv}a) confirm that the self-lensing signal in the \kepler\ light curve is due to a companion on a low-eccentricity, $450\mathchar`-\mathrm{day}$ edge-on orbit (Section \ref{ssec:model_rv}). Combined with the primary mass of $\approx1.1\,\msun$ from the Subaru spectrum (Section \ref{ssec:model_spec}, Figure \ref{fig:spectrum}), the RV amplitude derives the companion mass of $0.2\,\msun$. In addition, the self-lensing signal shows that the companion is a compact object smaller than $\sim0.02\,\rsun$ (Section \ref{ssec:model_joint}, Figure \ref{fig:lcrv}b--c). Therefore, we have a strong case that the companion is an ELM WD, even though its light is not detected. The upper limit on the WD luminosity placed by the null detection of the secondary eclipse is consistent with the inferred age of the primary star and the WD cooling model.

\begin{deluxetable*}{lcccc}[!ht]
\tablecaption{Parameters of the KIC 8145411 System.\label{tab:params}}
\tablehead{
\colhead{} & \colhead{Spectrum} & \colhead{RV} & \colhead{RV  \& \kepler\ SAP} & \colhead{RV  \& \kepler\ PDC}
}
\startdata
(\textit{Orbital Parameters})\\
time of periastron passage &\nodata & $3233^{+11}_{-12}$ & $488.6^{+3.7}_{-6.5}$ & $489.6^{+3.1}_{-4.5}$\\
\quad $\tau$ ($\mathrm{BJD}_\mathrm{TDB}-2454833$)\\
period $P$ (days) &\nodata& $448.6^{+5.8}_{-5.2}$ & $455.870^{+0.015}_{-0.015}$ & $455.826^{+0.009}_{-0.011}$\\
$\sqrt{e}\cos\omega$ &\nodata& $0.019^{+0.063}_{-0.066}$ & $-0.024^{+0.018}_{-0.016}$ & $-0.022^{+0.014}_{-0.014}$\\
$\sqrt{e}\sin\omega$ &\nodata& $-0.364^{+0.028}_{-0.021}$ & $-0.363^{+0.030}_{-0.021}$ & $-0.378^{+0.017}_{-0.019}$\\
RV semi-amplitude $K$ (km/s) &\nodata& $4.73^{+0.14}_{-0.13}$ & $4.579^{+0.083}_{-0.102}$ & $4.623^{+0.077}_{-0.075}$\\
RV zero-point $\gamma$ (m/s) &\nodata& $-687^{+75}_{-74}$ & $-748^{+65}_{-68}$ & $-726^{+62}_{-58}$\\
\vspace{0.2cm}
constant acceleration $\dot\gamma$ (m/s/day) &\nodata& $-0.33^{+0.31}_{-0.32}$ & $-0.52^{+0.32}_{-0.30}$ & $-0.51^{+0.31}_{-0.28}$\\
semi-major axis $a$ (au) &\nodata& $1.254^{+0.031}_{-0.031}$\tablenotemark{$\dagger$} & $1.270^{+0.027}_{-0.028}$ & $1.276^{+0.027}_{-0.028}$\\
eccentricity $e$	 &\nodata& $0.135\pm0.018$ & $0.132^{+0.016}_{-0.021}$ & $0.143^{+0.015}_{-0.012}$\\
argument of periastron $\omega$ (degrees) &\nodata & $-86.9^{+9.6}_{-10.7}$ & $-93.7^{+2.8}_{-2.5}$ & $-93.4^{+2.1}_{-2.2}$\\
inclination $i$ (degrees) &\nodata & \nodata & $89.929^{+0.046}_{-0.035}$ & $89.976^{+0.017}_{-0.025}$\\
periastron distance $a(1-e)$ (au) &\nodata& $1.084^{+0.033}_{-0.033}$\tablenotemark{$\dagger$} & $1.102^{+0.036}_{-0.033}$ & $1.091^{+0.031}_{-0.032}$\\
mass function ($\msun$) &\nodata& $4.77^{+0.42}_{-0.38}\times10^{-3}$ & $4.41^{+0.24}_{-0.27}\times10^{-3}$ & $4.52^{+0.22}_{-0.21}\times10^{-3}$\\
\\
(\textit{Primary Parameters})\\
mass $M_1$ ($\msun$) & $1.11\pm0.08$ & \nodata & $1.118^{+0.079}_{-0.077}$ & $1.132^{+0.078}_{-0.078}$\\
radius $R_1$ ($\rsun$) & $1.27\pm0.18$ & \nodata & $1.195^{+0.082}_{-0.084}$ & $1.269^{+0.045}_{-0.047}$\\
effective temperature $T_\mathrm{eff}$ (K) & $5724\pm110$ & \nodata & \nodata & \nodata\\
surface gravity $\log_{10}\,(g/\mathrm{cm\,s^{-2}})$ & $4.20\pm0.12$ & \nodata & \nodata & \nodata\\
metallicity $\mathrm{[Fe/H]}$ & $0.39\pm0.09$ & \nodata & \nodata & \nodata\\
age $\log_{10}\,(\mathrm{age}/\mathrm{yr})$ & $9.81\pm0.17$ & \nodata & \nodata & \nodata\\
\\
(\textit{White-dwarf Parameters})\\
mass $M_2$ ($\msun$) &\nodata&  $0.201^{+0.012}_{-0.011}$\tablenotemark{$\dagger$} & $0.197^{+0.010}_{-0.010}$ & $0.200^{+0.009}_{-0.009}$\\
radius $R_2$ ($\rsun$) &\nodata&\nodata&$<0.022$ ($99.7\%$ limit)&$<0.015$ ($99.7\%$ limit)\\
\\
(\textit{Light-curve Parameters})\\
time of inferior conjunction &\nodata & $313^{+37}_{-41}$ & $267.895^{+0.016}_{-0.017}$ & $267.867^{+0.022}_{-0.017}$\\
\quad $t_{\rm 0,pulse}$ ($\mathrm{BJD}_\mathrm{TDB}-2454833$)\\
time of superior conjunction  &\nodata & $539^{+31}_{-34}$ & $493.3^{+1.9}_{-1.7}$ & $493.4^{+1.5}_{-1.6}$\\
\quad $t_{\rm 0,secondary}$ ($\mathrm{BJD}_\mathrm{TDB}-2454833$)\\
impact parameter (primary eclipse) $b_{\rm pulse}$ &\nodata& \nodata& $0.32^{+0.15}_{-0.20}$ & $0.10^{+0.11}_{-0.07}$\\
impact parameter (secondary eclipse) $b_{\rm secondary}$ &\nodata& \nodata& $0.25^{+0.11}_{-0.16}$ & $0.078^{+0.079}_{-0.055}$\\
limb-darkening coefficient $q_1=(u_1+u_2)^2$ & \nodata & \nodata & $0.68^{+0.23}_{-0.30}$ & $0.38^{+0.34}_{-0.23}$\\
limb-darkening coefficient $q_2=u_1/2(u_1+u_2)$ & \nodata & \nodata & $0.54^{+0.29}_{-0.32}$ & $0.37^{+0.35}_{-0.25}$\\
flux normalization (1st pulse) $c_1$ & \nodata & \nodata & $1.000001^{+0.000021}_{-0.000020}$ & $1.000004^{+0.000021}_{-0.000020}$\\
flux normalization (2nd pulse) $c_2$ & \nodata & \nodata & $1.000033^{+0.000026}_{-0.000026}$ & $1.000023^{+0.000020}_{-0.000020}$\\
\\
(\textit{Jitters})\\
RV jitter $\log_{10}\sigma_{\rm RV}$ (m/s) & \nodata& $1.95^{+0.32}_{-1.02}$ & $2.13^{+0.22}_{-0.40}$ & $2.06^{+0.24}_{-0.55}$\\
photometric jitter (1st pulse) $\log_{10}\sigma_1$ &\nodata& \nodata & $<-3.9$ ($99.7\%$ limit) & $<-3.9$ ($99.7\%$ limit)\\
photometric jitter (2nd pulse) $\log_{10}\sigma_2$ &\nodata& \nodata & $-3.672^{+0.058}_{-0.069}$ & $<-3.8$ ($99.7\%$ limit)\\
\enddata
\tablenotetext{\dagger}{These values are derived assuming $i=90^\circ$ and the primary mass from spectroscopy.}
\tablecomments{The values are the medians and symmetric $68.3\%$ regions of the marginal posteriors.}
\end{deluxetable*}

\subsection{Characterization of the Primary Star}\label{ssec:model_spec}

\begin{figure*}[!ht]
	\epsscale{0.9}
	\plotone{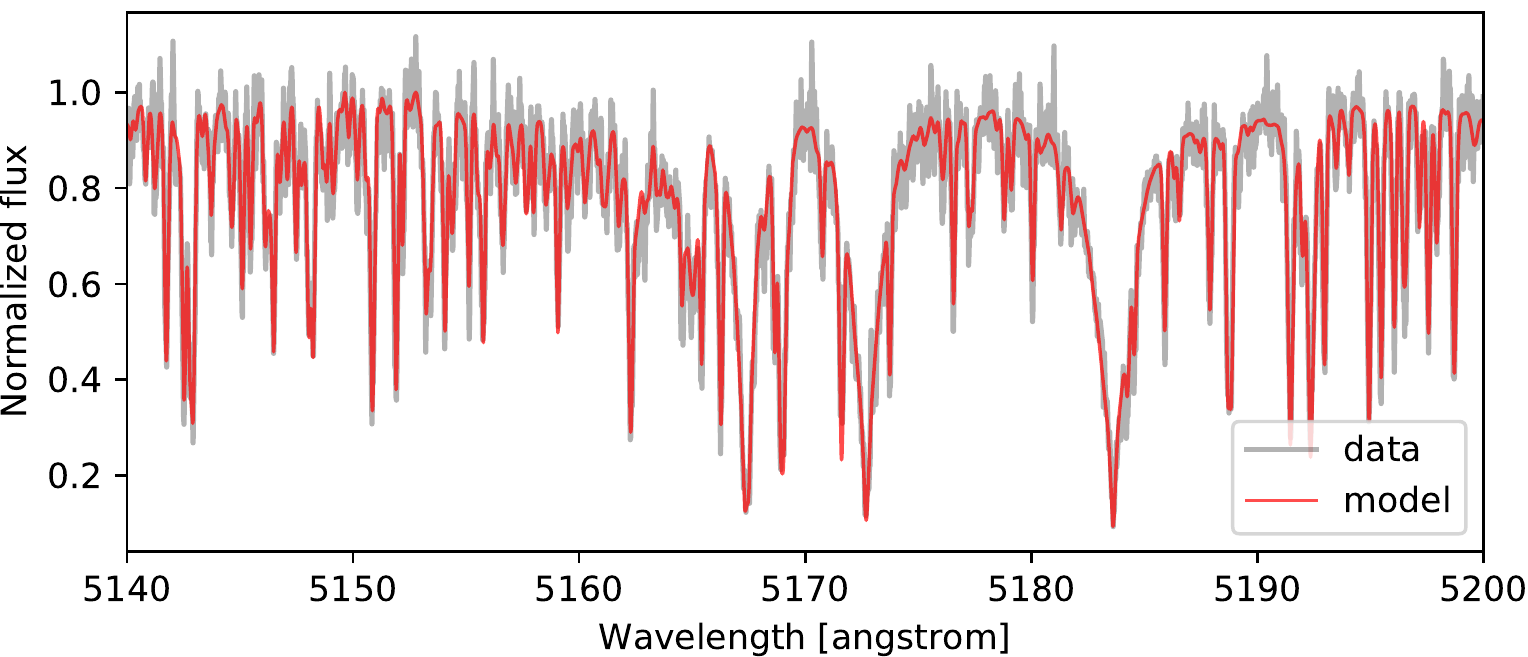}
	\caption{A portion of the HDS spectrum of the primary star (gray) and the best-match model from {\tt SpecMatch-Emp} (red).
	}
	\label{fig:spectrum}
\end{figure*}

We characterized the visible primary star KIC 8145411 applying the {\tt SpecMatch-Emp} code \citep{2017ApJ...836...77Y} to its high-resolution spectrum (Section \ref{ssec:data_hds}). 
The stellar parameters are estimated by matching the observed spectrum against a library of high-resolution ($R\approx60,000$), high signal-to-noise ($\approx150$ per pixel) Keck/HIRES spectra of $\sim$M5--F1 touchtone stars with well-determined physical properties. The input spectrum was resampled and shifted onto the library wavelength scale using cross-correlation, and the best-match library spectra that minimize the sum of squares of the difference were searched by applying a rotational broadening kernel to the library spectra and adjusting $v\sin i$ along with the continuum normalization. Five best-matching library spectra and physical parameters of those stars were identified and linearly interpolated to find the best-match composite spectrum (Figure \ref{fig:spectrum}) and the corresponding set of parameters of the target star. The uncertainties of the parameters are estimated based on the accuracy of the parameter recovery when the code is applied to their library stars with known parameters. Since the parallax measurement for KIC 8145411 is not available in the Data Release 2 of \gaia\ \citep{2018A&A...616A...1G}, we simply adopted the outputs of the code, as summarized in the ``Spectrum" column in Table \ref{tab:params}. Although the covariance between the parameters is not taken into account, that does not affect our analysis because their correlation is dominated by the constraint on the mean primary density from self-lensing duration, which scales as $P^{1/3}(M_1/R_1^3)^{-1/3}$ \citep[e.g.,][]{2011exop.book...55W}.

Considering the importance of the stellar parameter estimation, we also applied the {\tt ZASPE} code \citep{2017MNRAS.467..971B} to the same Subaru spectrum. We found $T_{\rm eff}=5751\pm96\,\mathrm{K}$,  $\log g=4.34\pm0.20$, $\mathrm{[Fe/H]}=0.36\pm0.05$, and $v\sin i=5.32\pm0.25\,\mathrm{km/s}$, which are all consistent with the results from {\tt SpecMatch-Emp}.\footnote{Although $v\sin i$ is not included in the outputs of the {\tt SpecMatch-Emp} code, the values used in the five best-match spectra are consistent with the estimate from {\tt ZASPE}.} The resulting effective temperature is also consistent with the value ($5713^{+154}_{-188}\,\mathrm{K}$) in the \kepler\ input catalog \citep{2017ApJS..229...30M}, which is essentially based on broad-band photometry.

\subsection{Spectroscopic Orbit}\label{ssec:model_rv}

We fitted a Keplerian model to the TRES RV time series (Table \ref{tab:rvdata}, Section \ref{ssec:data_rv}) to determine the orbital period $P$, RV semi-amplitude $K$,  eccentricity and argument of periastron referred to the sky $\sqrt{e}\cos\omega$ and $\sqrt{e}\sin\omega$, time of periastron passage $\tau$, radial-velocity zero point $\gamma$, constant acceleration $\dot\gamma$, and the RV jitter $\sigma_{\rm RV}$. 
We used {\tt emcee} \citep{2013PASP..125..306F} to sample from the posterior distribution of these parameters via a Markov chain Monte Carlo algorithm with the affine-invariant ensemble sampler \citep{2010CAMCS...5...65G}.
The log-likelihood function was defined as
\begin{equation}
	\label{eq:lnlike}
	\ln\mathcal{L}=-{1\over2}\sum_i {(y_i-y_{{\rm model},i})^2\over{\sigma_i^2+\sigma_{\rm RV}^2}}-{1\over2}\sum_i \ln\,(\sigma_i^2+\sigma_{\rm RV}^2),
\end{equation}
where $y$ and $\sigma$ denote the observed RV value and its error, respectively, and $y_{\rm model}$ is the modeled RV value.
We adopted uniform priors for $\tau$, $\sqrt{e}\cos\omega$, $\sqrt{e}\sin\omega$, $\gamma$, and $\dot\gamma$; and log-uniform ones for $P$, $K$, and $\sigma_{\rm RV}$. 

The resulting constraints are summarized in the ``RV" column in Table \ref{tab:params}. The orbital period turned out to be half the interval of the two detected self-lensing pulses, meaning that one pulse was missed in the data gap. The binary mass function $(4.8\pm0.4)\times10^{-3}\,M_\sun$, combined with the primary mass $1.12\pm0.08\,\msun$ (Section \ref{ssec:model_spec}) and the orbital inclination $\approx90^\circ$ (as justified from self-lensing signals), yields the WD mass of $0.20\pm0.01\,\msun$. We also find the orbital semi-major axis of $1.26\pm0.03\,\mathrm{au}$ and eccentricity of $0.14\pm0.02$.

We did not find strong evidence for radial acceleration $\dot\gamma$ due to a tertiary companion. The current limit $|\dot\gamma|\lesssim1\,\mathrm{m/s/day}$ implies $m_3\sin i_3/a_3^2\lesssim M_\odot/(20\,\au)^2$ for the mass $m_3$, orbital inclination $i_3$, and semi-major axis $a_3$ of the potential tertiary companion.

\subsection{Combined Analysis}\label{ssec:model_joint}

To check that the RV time series and the \kepler\ self-lensing light curves are consistently explained, we jointly modeled the two data sets incorporating the knowledge of the primary star from spectroscopy. We defined the likelihood function for the photometric data in a similar way to Equation \ref{eq:lnlike}, where the additional white noise term (photometric jitter) was introduced analogously to $\sigma_{\rm RV}$. Here we adopted independent photometric jitters for the first and second pulses because the \kepler\ data around the second pulse were obtained during reaction-wheel zero crossings and the raw SAP light curve shows some artificial dips that have been corrected in the PDC pipeline \citep{2012PASP..124..985S, 2012PASP..124.1000S}. The correction could introduce systematics that affect the noise level, and the two jitter terms are introduced to account for this possibility. 
The self-lensing light curves were modeled mostly in the same way as in \citet{2018AJ....155..144K}: the light curve was computed as the sum of the WD eclipse and self-lensing pulse, where the latter was approximated as an inverted eclipse whose height is determined by the Einstein radius of the WD instead of its physical radius \citep{2003ApJ...594..449A}. The normal/inverted eclipse light curves were computed using the \citet{2002ApJ...580L.171M} model for the quadratic limb-darkening law as implemented in {\tt PyTransit} \citep{2015MNRAS.450.3233P}, where the limb-darkening coefficients $(u_1, u_2)$ were parameterized following \citet{2013MNRAS.435.2152K}. Here we chose the WD radius to be a free parameter rather than a deterministic function of the WD mass: given that the amplitude of the self-lensing pulse is fixed through the WD mass from RVs, the WD radius can be constrained from the eclipse component \citep[cf.][]{2019arXiv190411063Y}.

The results are summarized in Table \ref{tab:params} and the posterior models are shown in Figure \ref{fig:lcrv} with gray lines. The RVs and light curves are consistently modeled, and the resulting constraints are mostly unchanged from the RV-only model except that the orbital period is more precisely determined. The main results including the WD mass are mostly insensitive to which of the PDC and SAP light curves is adopted, either. We confirmed the compact nature of the companion, placing upper limits on its radius, $<0.022\,\rsun$ from the SAP data and $<0.015\,\rsun$ from the PDC data, which are independent from the WD model. 

The zero-temperature WD model predicts $\approx0.02\,\rsun$ for the measured mass of $0.2\,\msun$ (\citealt{1972ApJ...175..417N}; see also figure 13 of \citealt{2018AJ....155..144K}), which is formally inconsistent with the PDC value and barely consistent with the SAP one. Although these limits, if taken at face value, are potentially interesting, we note that the quantitative constraint on this parameter depends on subtle eclipse signals of order $10^{-4}$ and would be of limited reliability given the quality issue of the light curve discussed above. Indeed, comparing the data and models in Figure \ref{fig:lcrv} b and c, the second pulse appears to be slightly higher; this suggests that unmodeled systematics exist in the data, which may also be the origin of different limits obtained from SAP and PDC data. 
Thus at the moment we do not consider this limit to be sufficiently strong evidence for more exotic explanations including a black hole and a neutron star, although the current data, including the absence of secondary eclipses as discussed below, are compatible with such scenarios as well.
Follow-up observations of more self-lensing events with a precision comparable to or better than \kepler\ would be required to resolve this issue.
To make sure that this possible tension does not significantly affect the measured WD mass, we also repeated the same analysis for the PDC data imposing the Eggleton mass--radius relation \citep{1988ApJ...332..193V} for the companion. We found $0.22\pm0.01\,\msun$ for the WD mass, which is only marginally different from the above result. The best-fit model from this analysis is shown with red dashed lines in Figure \ref{fig:lcrv}. The small difference from the previous analysis (gray lines) highlights the subtlety involved in measuring the WD radius accurately.

\begin{figure}[!ht]
	\epsscale{1.08}
	\plotone{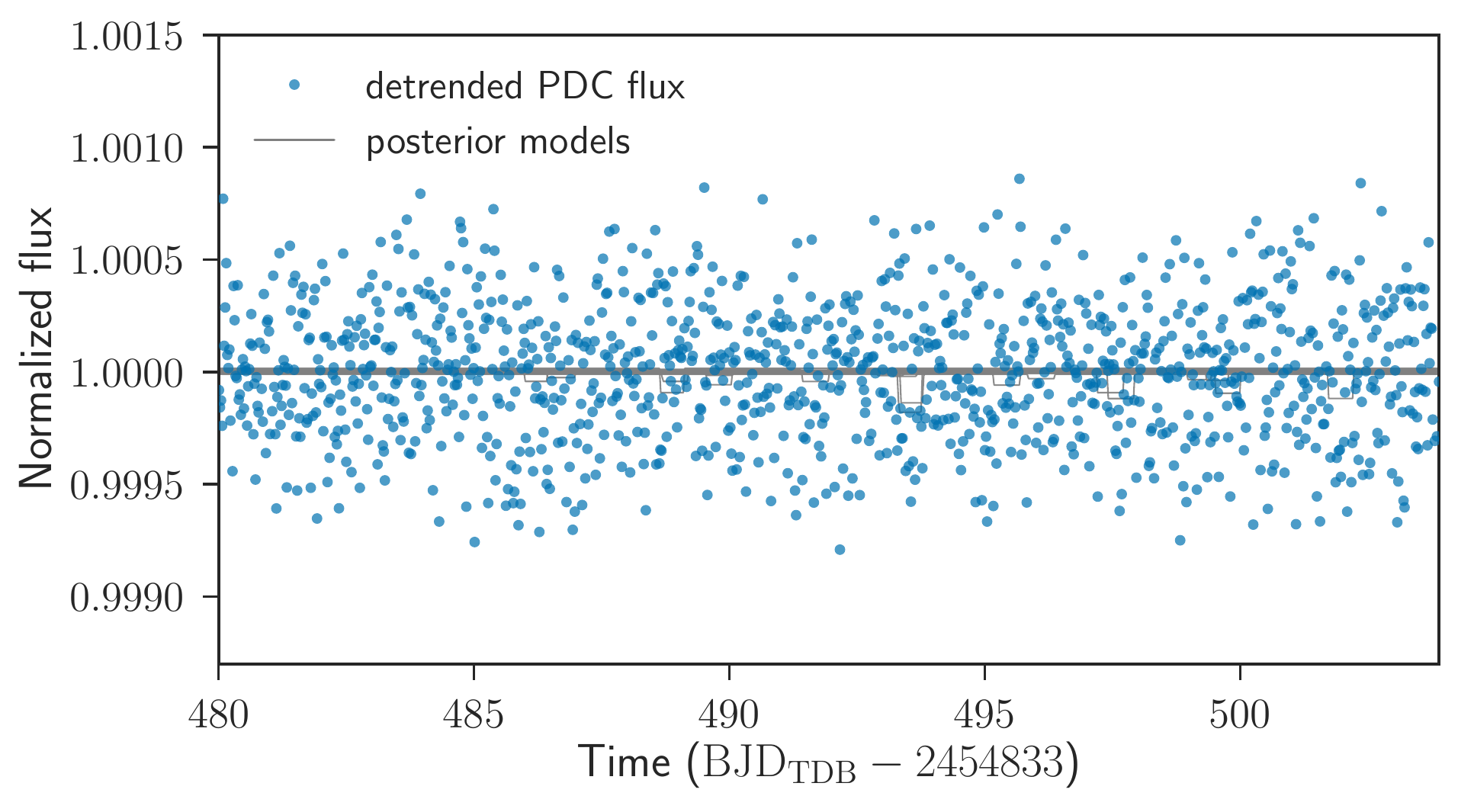}
	\epsscale{1.1}
	\plotone{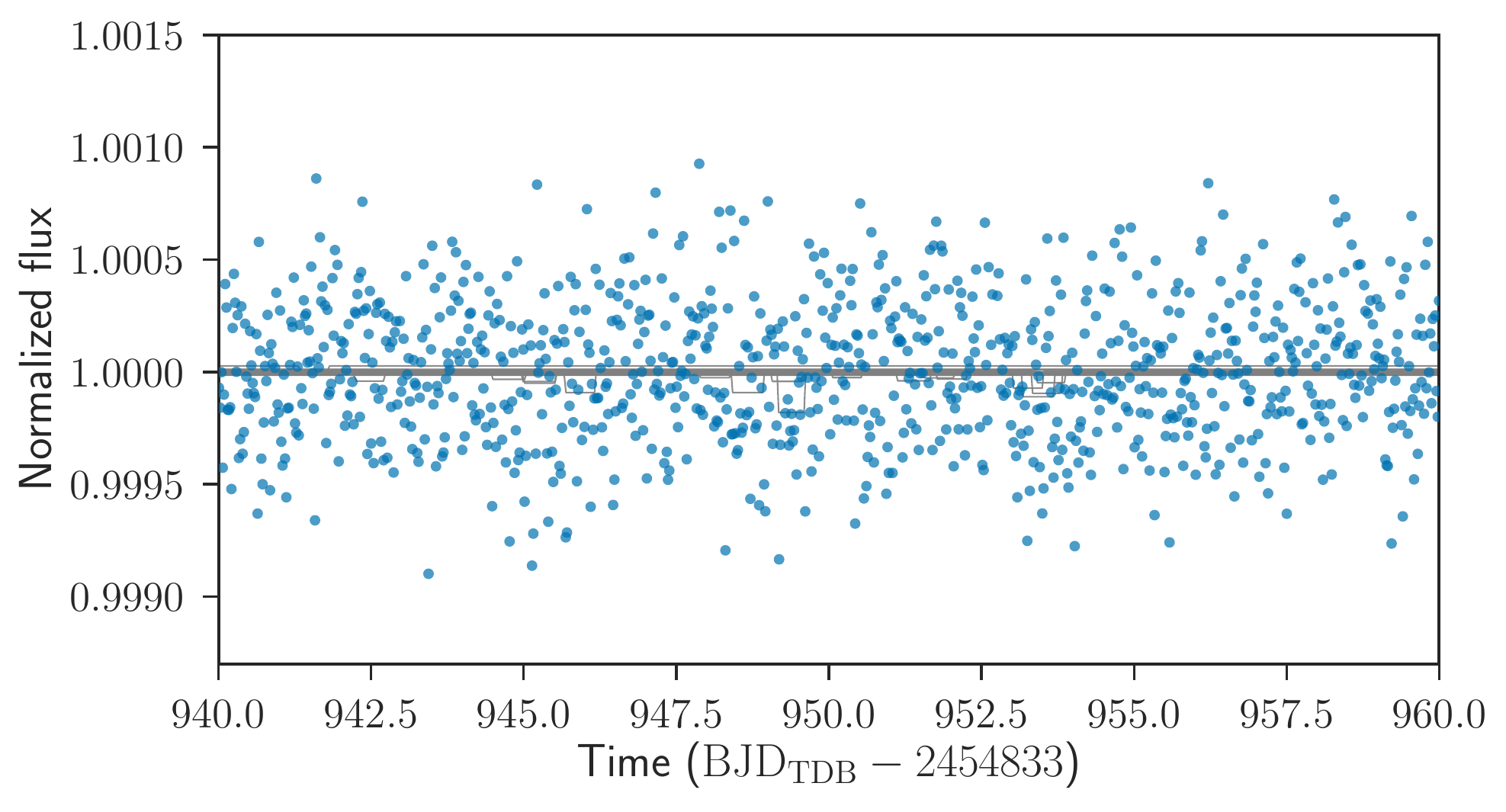}
	\epsscale{1.1}
	\plotone{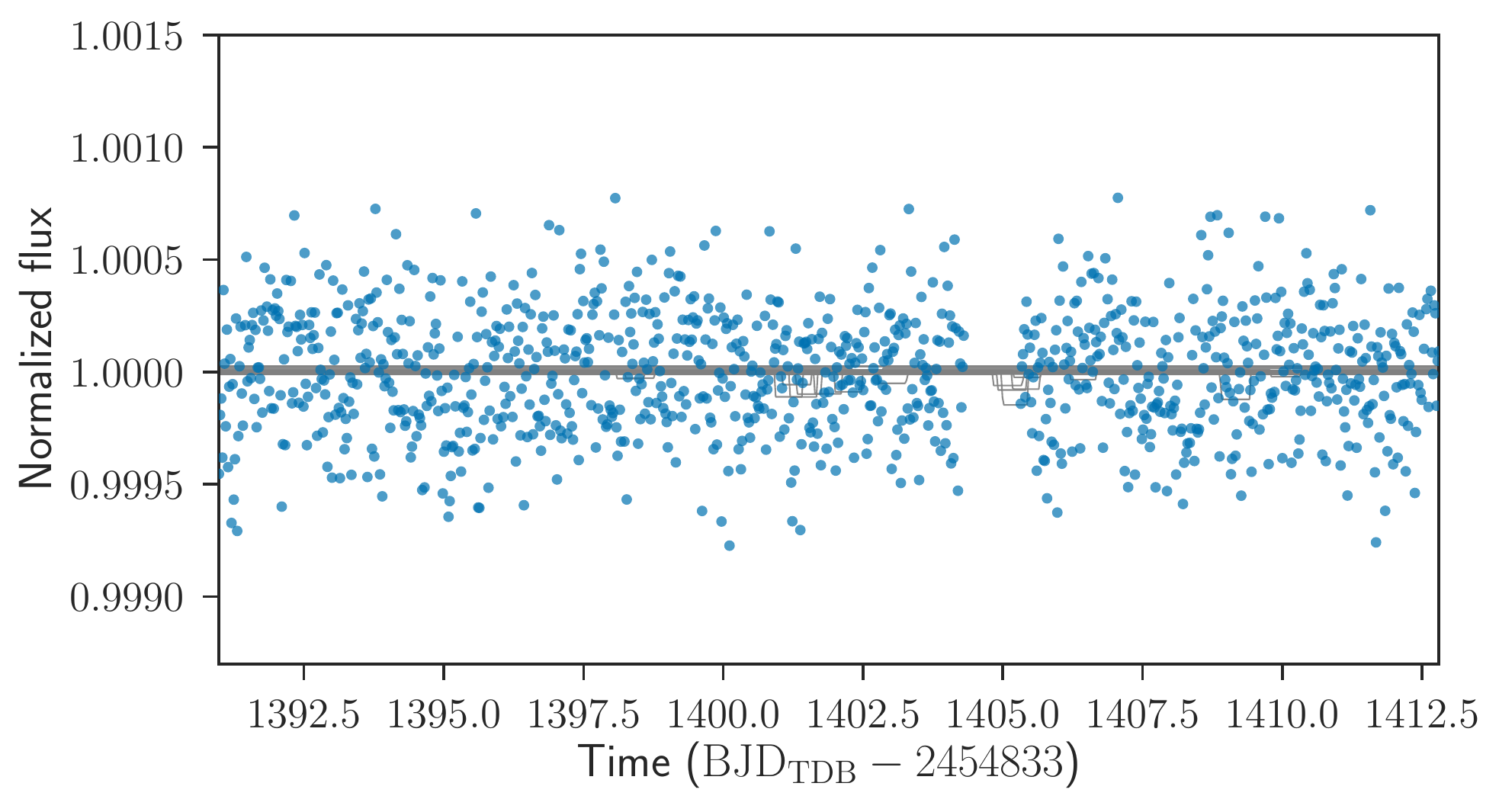}
	\caption{The detrended PDC light curves around the expected times of secondary eclipses (blue points) and random 20 posterior models (gray lines) from the joint analysis of RVs, self-lensing pulses, and the data shown in this plot (Section \ref{ssec:model_joint}). Here the WD radius is related to its mass via a physical mass--radius relation and the effective temperature of the WD is an additional free parameter.
	}
	\label{fig:secondary}
\end{figure}

\begin{figure}[!ht]
	\epsscale{1.15}
	\plotone{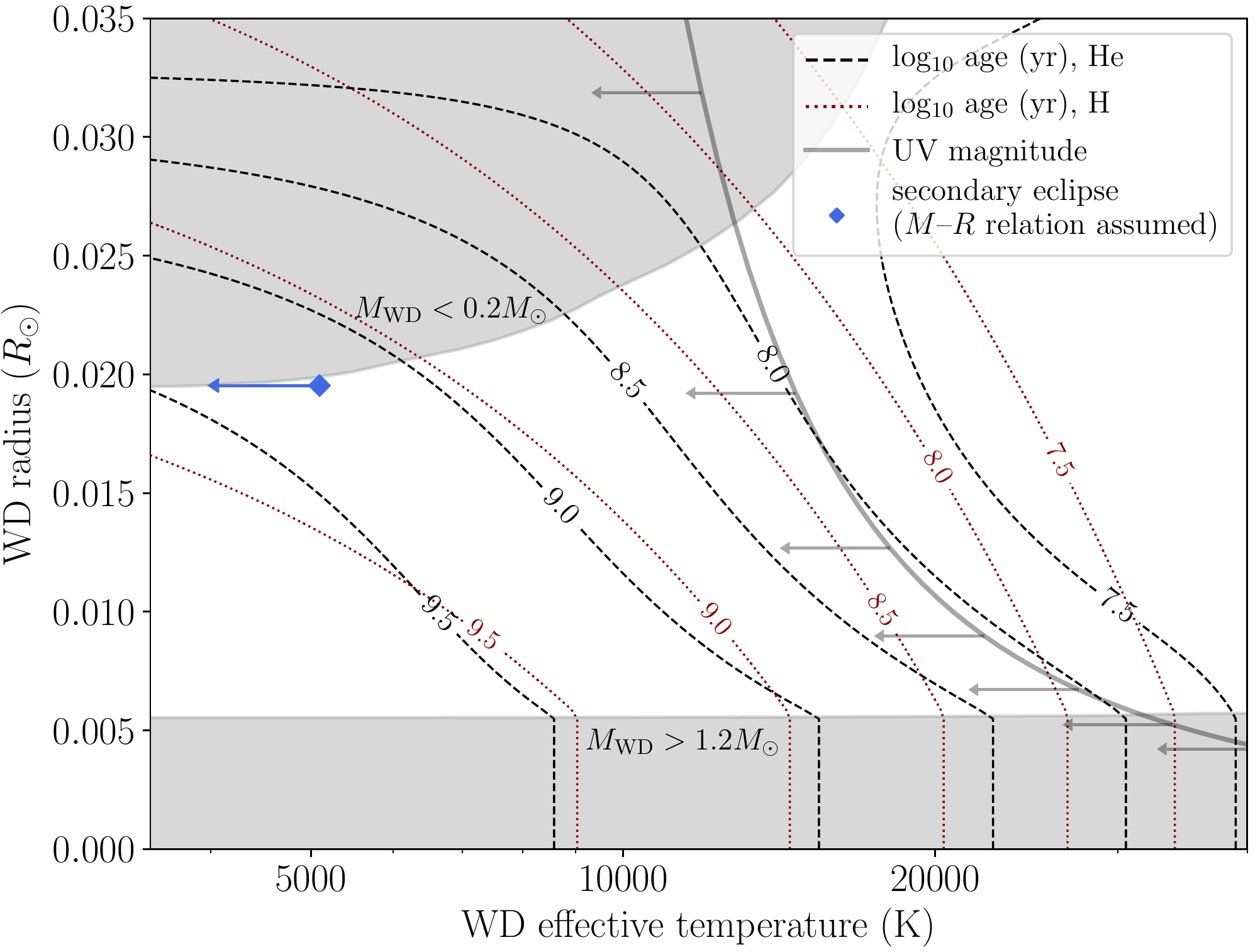}
	\caption{Constraints on the WD radius and effective temperature based on the absence of secondary eclipses and physical mass--radius relation (blue diamond), compared with evolutionary models for hydrogen- (red dotted) and helium- (black dashed) atmosphere WDs from \url{http://www.astro.umontreal.ca/~bergeron/CoolingModels} \citep{2006AJ....132.1221H, 2006ApJ...651L.137K, 2011ApJ...730..128T, 2011ApJ...737...28B}. The models are not available for gray-shaded regions. The gray solid line shows the constraint from the absence of UV excess in the {\it GALEX} data; see Section 3.3 of \citet{2018AJ....155..144K} for details.
	}
	\label{fig:cooling}
\end{figure}

The impact parameter during self-lensing events, along with $e$ and $\omega$, indicates that the WD was totally occulted by the primary. Nevertheless, we did not clearly detect secondary eclipses at the expected times. 
To check if this is consistent with the WD interpretation, we performed another analysis including the PDC data around the expected secondary eclipse times and imposing the Eggleton mass--radius relation. Here the WD effective temperature $T_{\rm WD}$ is an additional free parameter constrained from the absence of secondary eclipses, whose depth in the model was computed by convolving the blackbody spectra of the WD and the primary star with the response function of \kepler. The effective temperature of the primary was fixed to be $5724\,\mathrm{K}$ from spectroscopy. The posterior models around expected secondary eclipse times from this analysis are compared with the data in Figure \ref{fig:secondary}. The lack of clear secondary eclipses requires $T_{\rm WD}\lesssim5000\,\mathrm{K}$ ($95\%$ limit), which is consistent with theoretical evolutionary models of WDs older than a few Gyr (Figure \ref{fig:cooling}). 
This is also compatible with the crude spectroscopic age of the primary star $6.5^{+3.1}_{-2.1}\,\mathrm{Gyr}$, although the value is not very well constrained and may need to be interpreted with care if the system has experienced mass transfer in the past.
Considering that the system has been identified via self-lensing, it is not very unnatural that the detected WD has cooled down because self-lensing signals are diminished for younger low-mass WDs with larger physical radii.

\section{``Impossible" ELM WD?}\label{sec:impossible}

\begin{figure*}
	\epsscale{1.18}
	\plotone{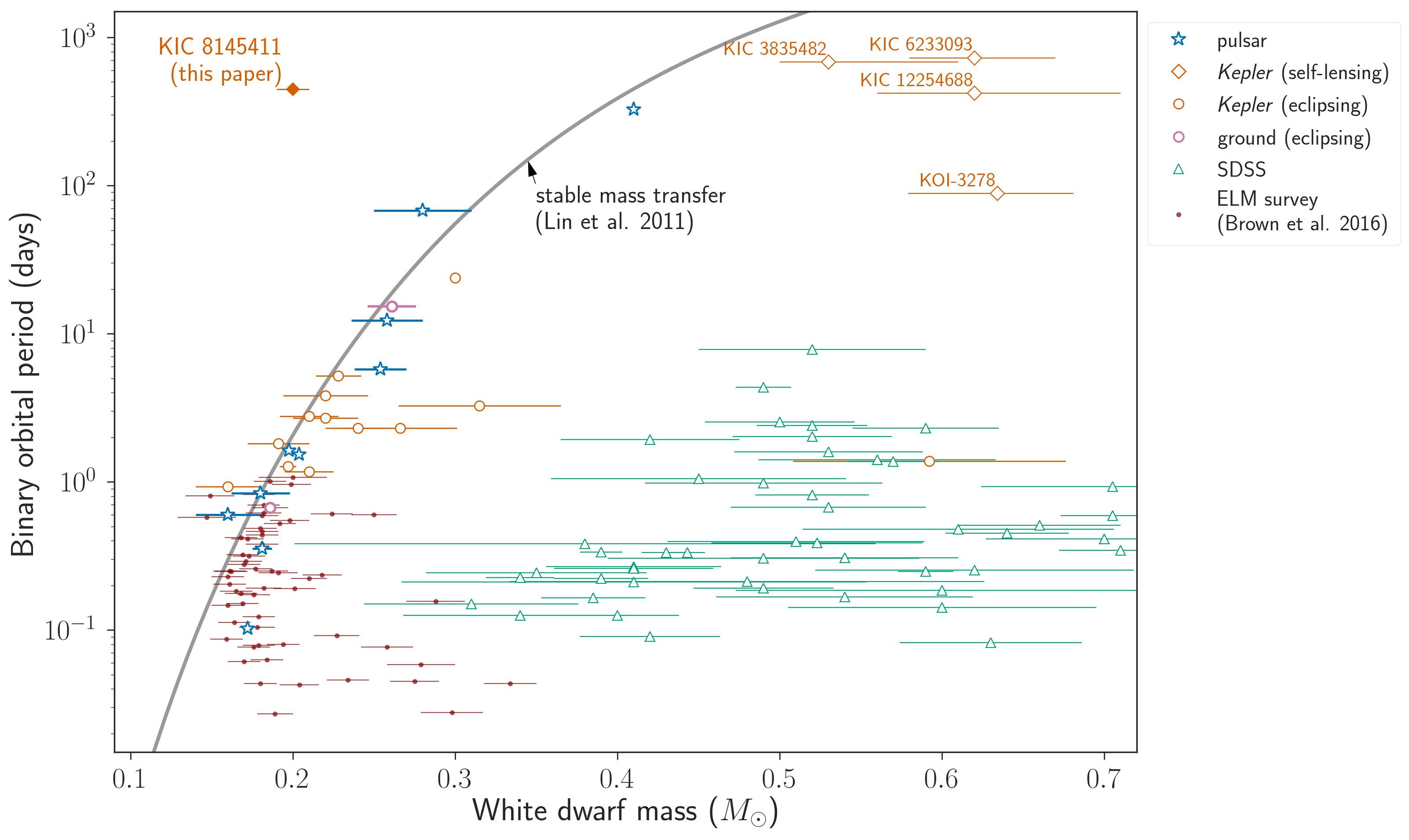}
	\caption{The masses of WDs in binaries and their orbital periods. The KIC 8145411 system (upper left) and other self-lensing systems \citep{2014Sci...344..275K, 2018AJ....155..144K} are shown with orange diamonds. The gray line is the theoretical $P$--$M_{\rm WD}$ relation for the stable mass transfer case by \citet{2011ApJ...732...70L}. Blue stars denote the systems with pulsar companions compiled by \citet{2014ApJ...781L..13T}; orange and pink circles are those eclipsing stellar companions from \kepler\ \citep{2010ApJ...715...51V, 2011ApJ...728..139C, 2012ApJ...748..115B, 2015ApJ...803...82R, 2015ApJ...815...26F, 2017ApJ...850..125Z, 2013ApJ...767..111M} and ground-based data \citep[OGLE and WASP surveys;][]{2012Natur.484...75P, 2013Natur.498..463M}, respectively; green triangles are WD--main-sequence binaries from the SDSS \citep{2012MNRAS.419..806R}; filled maroon circles are from the ELM survey \citep{2016ApJ...818..155B}. 
	}
	\label{fig:pmplane}
\end{figure*}

The orbit of the KIC 8145411 system is far wider than known binaries containing ELM WDs and challenges the standard formation scenario through RGB mass transfer. To produce such a WD, 
the envelope of the WD progenitor needs to be stripped
when its core had $\approx0.2\,\msun$ in the RGB phase. The radius of such an RGB star never exceeds $9\,\rsun\approx0.04\,\au$, which we confirmed using the {\tt MESA} code \citep{2011ApJS..192....3P, 2013ApJS..208....4P, 2015ApJS..220...15P, 2018ApJS..234...34P}, and the orbit needs to be comparably tight for the binary to interact. Indeed, known binaries with ELM WDs have such tight orbits as shown in Figure \ref{fig:pmplane}, where we plot the WDs in binaries and with measured masses collected from the literature.
The gray line quantifies the above argument about the relation between the orbit and the WD mass with numerical calculations for the stable mass transfer case \citep{2011ApJ...732...70L}: this well matches the upper envelope of the known systems, regardless of the WD mass and type of companions. Thus they are compatible with formation through stable mass transfer, or common-envelope evolution in which the orbit shrank due to dynamical friction \citep{1976IAUS...73...75P}. However, the KIC 8145411 system has a much wider orbit than predicted from this relation, and it should have been impossible for the progenitor of the WD, when its core was $\approx0.2\,\msun$, to have filled its Roche lobe. Considering that the progenitor was initially more massive, the Roche lobe radius around the progenitor was $\gtrsim0.5\,\au$ for $a=1.28\,\au$, which is $>10$ times larger than the radius of the progenitor.

We do not have a definitive solution to this puzzle, and there appears to be something missing in our understanding of the formation of ELM WDs and/or binary interactions. In the following sections, we discuss observational implications of this finding, and briefly comment on possible formation paths.

\subsection{Could the Primary Mass be Incorrect?}

Could the anomalous WD mass be due to misclassification of the primary star? One might be worried that some peculiar evolutionary history of the system (e.g., significant mass transfer) could affect the interpretation of the spectrum. While we believe that parameters such as $T_{\rm eff}$, $\log g$, and [Fe/H] are unaffected because they are only based on model stellar atmospheres, it may still be possible that the relation between the atmospheric parameters and bulk stellar properties is systematically biased because they are tied using evolutionary models of a single isolated star. 

\begin{figure}
	\epsscale{1.15}
	\plotone{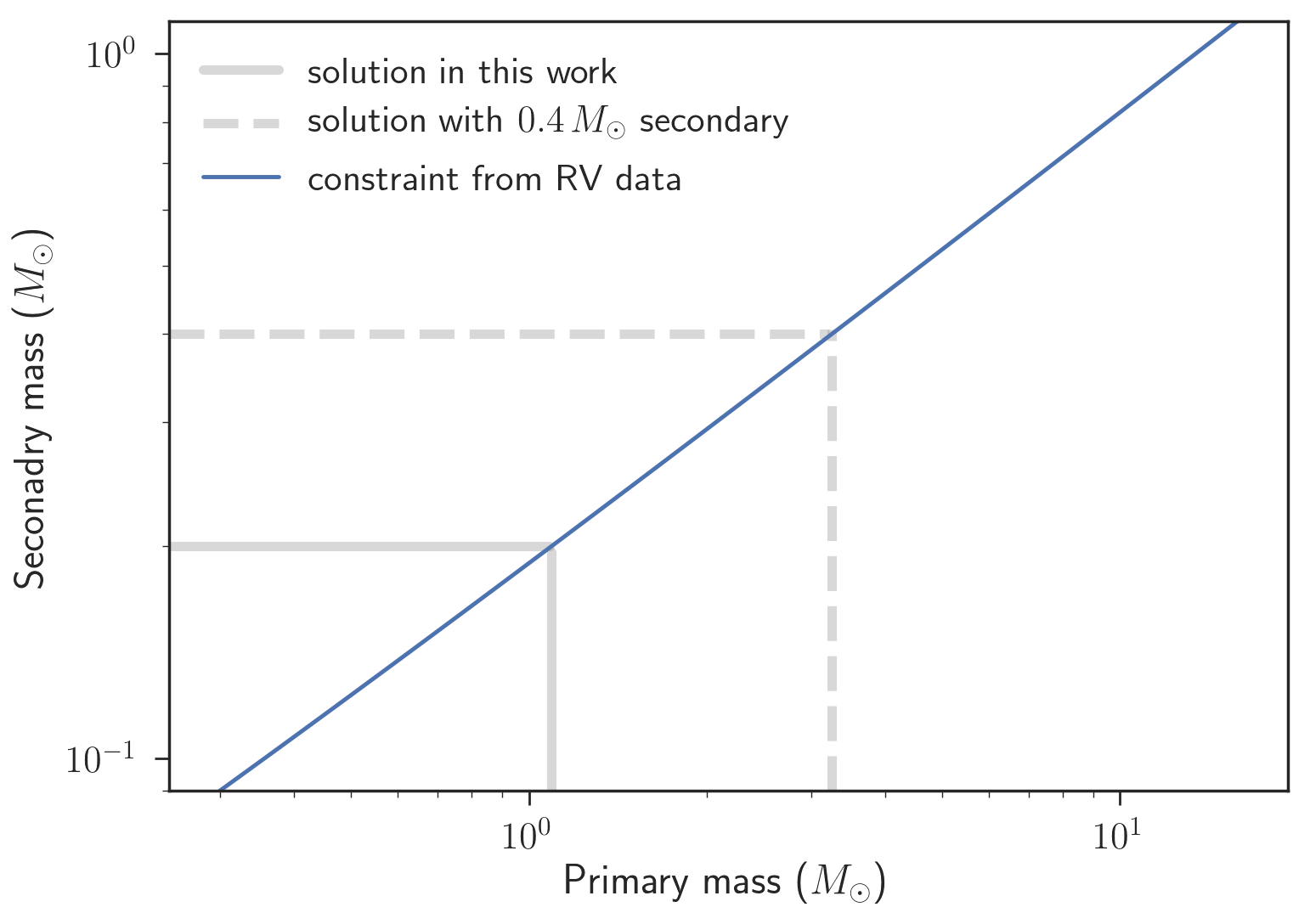}
	\caption{The secondary and primary masses from the RV data. The blue solid line shows their relation based on the observed binary mass function of $4.8\times10^{-3}\,M_\odot$. The thick solid and dashed gray lines show the solution found in this work (i.e., $0.2\,\msun$ companion) and that with a $0.4\,\msun$ companion as theoretically expected from the binary orbital period (gray line in Figure \ref{fig:pmplane}; see also Section \ref{sec:impossible}), respectively.
	}
	\label{fig:m2_m1}
\end{figure}

Given the difficulty of modifying stellar models accordingly, here we ask the following question: how wrong do the primary mass and radius need to be for the companion WD to be a ``normal" one? Figure \ref{fig:m2_m1} shows the relation between the secondary and primary masses for the binary mass function $4.8\times10^{-3}\,\msun$ {\it measured} from the RV data. The thick solid gray line shows that the primary mass of $\approx1.1\,\msun$ corresponds to the companion mass of $\approx0.2\,\msun$, as we presented. The thick dashed line shows that, if the companion has $\approx0.4\,\msun$ as theoretically expected from the observed binary period (see the gray line in Figure \ref{fig:pmplane}), the primary has to have $\approx3.3\,\msun$ to reproduce the observed RV amplitude. On the other hand, this change should also affect the amplitude of the self-lensing signal, which scales as $M_{\rm WD}/R_1^2$: thus, $R_1$ also needs to be $\sim\sqrt{2}$ times larger than the current estimate and should be $\sim1.8\,\rsun$ to produce the same self-lensing signal for $M_{\rm WD}=0.4\,\msun$. Therefore, the current RV and self-lensing data allow the WD companion to have theoretically expected mass and orbital period only if the primary is a B- or A-type main-sequence star. This case is excluded from the spectrum presented in Section \ref{ssec:model_spec} and Figure \ref{fig:spectrum}.

\section{Occurrence Rate of Similar Systems and Connections to FBS Binaries}\label{sec:fbs}

The eclipse probability of the KIC 8145411 system is $R_1/a(1-e^2)\approx1/200$, while it was found among $\sim{10^5}$ \kepler\ stars showing sufficiently small photometric noise. Thus such ELM WD companions should occur around $\sim 200/10^5=0.2\%$ of Sun-like stars, and the actual rate could be higher if the search incompleteness is corrected. On the other hand, the occurrence rate of WD companions with au-scale orbits around Sun-like stars, regardless of the WD mass, is likely a few \% \citep[e.g.,][]{2017ApJS..230...15M, 2018MNRAS.474.4322M}.
Their ratio implies 
that $f_{\rm ELM}=0.2\%/\mathrm{a\ few\ \%}\sim10\%$ of WD companions of Sun-like stars on au-scale orbits are ELM WDs.
The estimate is also consistent, in the order-of-magnitude sense, with the fact that KIC 8145411 is one of four self-lensing WD binaries (i.e., $f_{\rm ELM}=25\%$) with similar periods identified in \citet{2018AJ....155..144K}.

\begin{figure*}
	\epsscale{1.18}
	\plotone{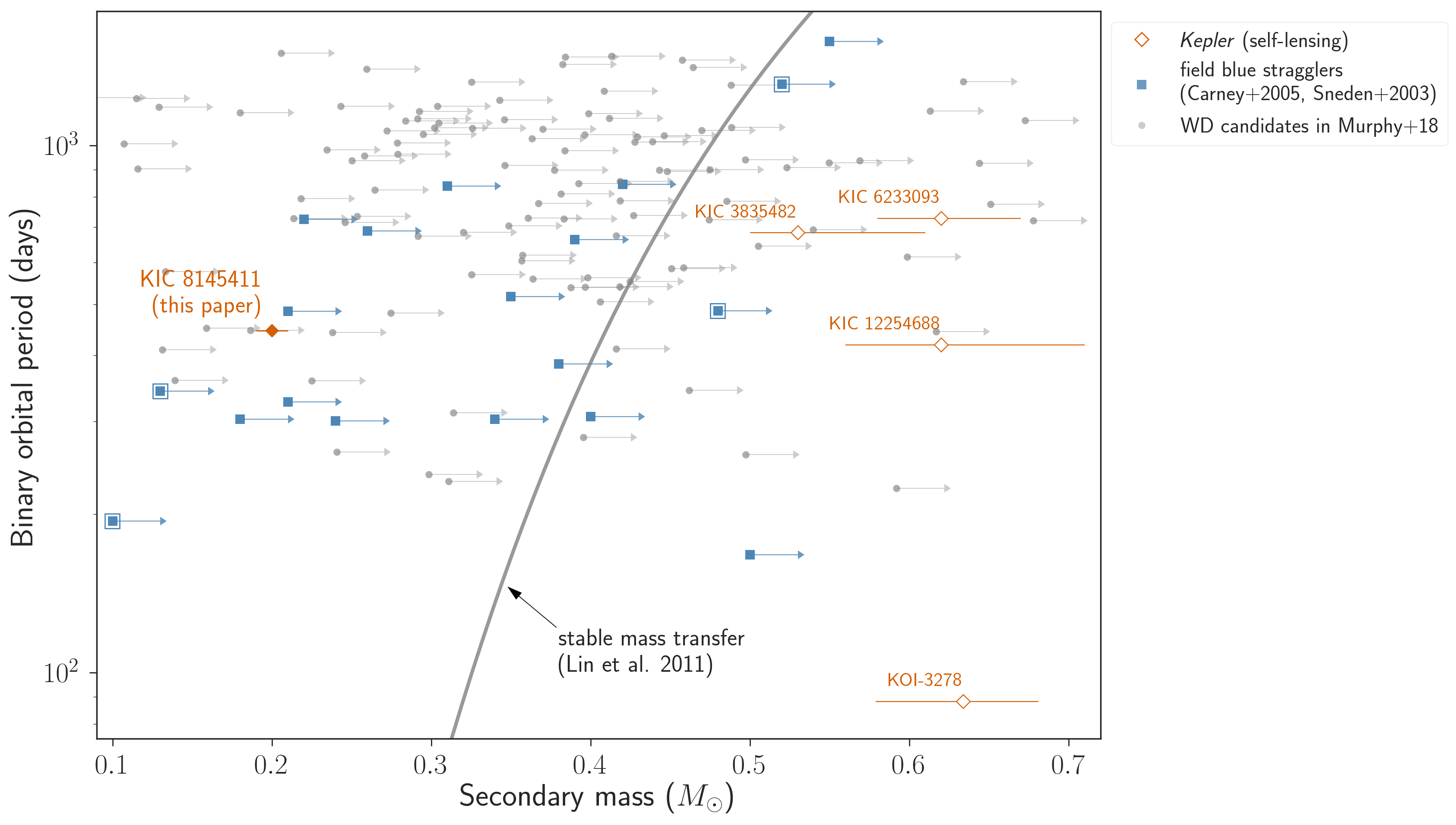}
	\caption{Same as Figure \ref{fig:pmplane}, but for WD candidates in stellar binaries with $P\gtrsim80\,\days$. Here the minimum masses of the companions in FBS binaries \citep{2005AJ....129..466C, 2003ApJ...592..504S} are shown by blue squares with arrows; those with outer squares are the systems with enhancements of neutron-capture elements detected \citep{2000AJ....120.1014P, 2003ApJ...592..504S}. Gray circles with arrows are the minimum masses of WD candidates identified in \citet{2018MNRAS.474.4322M}: $\approx 40\%$ of them are estimated to be WDs, although the identities of individual companions are unclear (see Section \ref{sec:fbs}).}
	\label{fig:pmplane2}
\end{figure*}

While the above $f_{\rm ELM}$ estimated from single detection is highly uncertain, the value may also be supported by a larger population of field blue straggler (FBS) binaries as studied in \citet{2000AJ....120.1014P, 2001AJ....122.3419C, 2005AJ....129..466C}, which carry many similarities to the KIC 8145411 system and other self-lensing WD binaries in \citet{2018AJ....155..144K}. Their primaries are old halo/thick-disk stars that appear to be too blue given their ages, and represent field analogs of blue stragglers in clusters. Their higher binary fraction, smaller companions masses, and lower eccentricities compared to normal binaries, absence of the companion's spectral features, and rapid rotation of the primary all suggest that a majority of these FBSs, if not all, are products of mass transfer \citep{1964MNRAS.128..147M} and that the unseen companions are WDs. 
Figure \ref{fig:pmplane2} compares these FBS binaries with self-lensing systems: the FBS sample is essentially the same as in Figure 10 of \citet{2005AJ....129..466C}, but here we computed the companion masses of CS 29497--030 and CS 29509--027 (not in Table 5 of their paper) using the RV solution in \citet{2003ApJ...592..504S} and the primary masses derived by fitting \citep{2015ascl.soft03010M} the Dartmouth isochrones \citep{2008ApJS..178...89D} to the atmospheric parameters in \citet{2014AJ....147..136R}, the parallax from \gaia\ DR2 \citep{2018A&A...616A...1G}, and the $K$-magnitude from the Two Micron All Sky Survey \citep[2MASS,][]{2006AJ....131.1163S}. Interestingly, some of the FBS companions are located close to KIC 8145411, 7 out of 19 having minimum masses $<0.25\,\msun$. For random orbital inclinations, this implies that effectively 4.6 would have {\it true} masses $<0.25\,\msun$ (i.e., $f_{\rm ELM}\approx20\%$). 

\citet{2018MNRAS.474.4322M} also identified similar systems with A/F-type primaries showing $\delta$-Scuti pulsations in the \kepler\ light curves. Analyzing pulsation-phase modulations caused by light-travel time effect, they identified non-eclipsing, unseen companions to these stars that contain an excess population of near-circular, au-scale binaries. \citet{2018MNRAS.474.4322M} argued that the companions in $\approx40\%$ of these $\approx120$ near-circular systems, defined by the relation $e<0.55\,\log(P/\mathrm{day})-1.21$ based on post-AGB binaries, are likely WDs. These near-circular systems are plotted with gray circles in Figure \ref{fig:pmplane2}, where the minimum secondary masses are computed assuming primary masses in the \kepler\ input catalog \citep{2017ApJS..229...30M}. 
If the sample indeed contains WDs and they have the same mass distribution as the whole sample, the minimum-mass distribution of the sample implies $f_{\rm ELM}\approx10\%$ for random orbital inclinations.

To summarize, in light of the detection in the KIC 8145411 system, it appears possible that some of the low-mass companions in the above two families of FBS binaries are also ELM WDs. If most of them are so, they imply $f_{\rm ELM}\approx10$--$20\,\%$, which supports the value inferred from the KIC 8145411 system alone.

This conjecture does not contradict the finding of the ELM survey that most of the surveyed ELM WDs were found to be in tight degenerate binaries \citep[e.g.,][]{2010ApJ...723.1072B}, because they are parts of the SDSS WDs, which do not include WDs with much brighter stellar companions like KIC 8145411. 
Rather, systems like KIC 8145411 could be progenitors of the ELM WDs in tight degenerate binaries: The mass transfer from the more massive, current stellar primary will likely be unstable, and the system may undergo common-envelope evolution to form a tight binary consisting of an ELM WD and a more massive WD.

\section{Summary and Discussion}\label{sec:discussion}

We confirmed the fifth self-lensing binary consisting of a $0.2\,\msun$ WD and a G-type star KIC 8145411.  The WD mass measured from RVs and self-lensing events is anomalously low given its au-scale orbit, which is far wider than required for the binary to have interacted to produce such a low-mass WD. Because only one in 200 such systems has edge-on geometry to show self-lensing, this system likely represents only the tip of the iceberg. More such low-mass WDs in non-eclipsing binaries may be identified with \gaia\ astrometry and/or RV search combined with complementary observations to show that the companion is too faint to be a star \citep[e.g.,][]{2019ApJ...875...74K}, providing a more complete view of post-interaction binaries with WDs and possibly the solution of the puzzle. Below we briefly comment on possible modifications to the standard binary evolution path to explain such anomalous systems, which all appear to involve some difficulties.

There may exist a path involving a tertiary star that made it possible for the binary to interact when the WD progenitor was in the RGB phase. For example, the orbit may have once been highly eccentric due to perturbations from the tertiary star, or may have been significantly widened via a rare stellar encounter after the binary interaction in a close-in orbit. However, it may be difficult for these scenarios to explain why the resulting orbit is nearly circular and why ELM WDs do not appear to be very rare among au-scale WD binaries, as argued in Section \ref{sec:fbs}.

Scenarios that do not involve currently detectable companions have also been proposed. \citet{1998A&A...335L..85N} proposed that the envelope of the WD progenitor could be stripped by a massive planet or a brown dwarf, which may then be merged with the resulting WD or evaporate to leave a single low-mass WD.\footnote{A similar path in stellar triple systems was also discussed by \citet{2019ApJ...876L..33P}, in which case the merged companion is assumed to be a star and significant mass ejection is required to leave a $0.2\,\msun$ white dwarf. The mass ejection tends to increase the orbital eccentricity, and so this scenario shares the same difficulty as the ones discussed here.}
\citet{2007ApJ...671..761K} argued that strong wind mass loss from metal-rich stars could truncate the AGB phase to produce single low-mass WDs. Although our system alone does not fully exclude these scenarios, they do not appear to be dominant channels because the low-eccentricity orbit of KIC 8145411, as well as those of the FBS binaries with ELM WDs (if they indeed are), suggest that generally the WD progenitors did interact with the observed stellar primaries. 

Alternatively, the WD progenitor might indeed have become large enough (e.g., an AGB star) for the binary to interact in the current $\sim1\,\au$ orbit. In this case, the core mass should have already been substantially larger than the current WD mass, as illustrated by the gray line in Figures \ref{fig:pmplane} and \ref{fig:pmplane2}, and the core somehow needs to lose $\sim50\%$ of its mass during or after the interaction. 
While we are not aware of such a mechanism, it is interesting to note that enhancements of carbon and neutron-capture elements have been detected in some of the FBS systems including those with smallest minimum-mass companions \citep[double squares in Figure \ref{fig:pmplane2};][]{2003ApJ...592..504S}, providing evidence for AGB mass transfer. Future measurements of true masses for these FBS companions with orbital inclinations from \gaia\ would further test this hypothesis.\\


\noindent {\it Note added in proof.} In the process of publishing this Letter we learned of a recent study by \citet{2018MNRAS.477L..40V}, who identified an ELM pre-WD (mass $0.23\pm0.05\,\msun$, effective temperature $26,200\pm1,500\,\mathrm{K}$, and log surface gravity $5.40\pm0.35$) in a 771-day period, near-circular orbit around a K-dwarf star. The hot primary HE 0430--2457 will evolve into a WD and occupy a similar region of the parameter space as the KIC 8145411 system in Figure \ref{fig:pmplane}. \citet{2018MNRAS.477L..40V} proposed a formation path similar to the one by \citet{1998A&A...335L..85N} discussed above, that the observed ELM pre-WD is a merger product of a tight binary that has lost a significant fraction of its mass, and that the current stellar companion was initially a tertiary star that did not directly interact with the pre-WD. However, this scenario still seems to have difficulty in explaining why the current orbit of the HE 0430--2457 system (as well as KIC 8145411) is nearly circular, as we discussed in Section \ref{sec:discussion}.

\acknowledgments

The authors thank Kenta Hotokezaka for stimulating discussions, and Simon Murphy for comments on the early manuscript. We also thank an anonymous referee for a prompt and insightful report that helped improving the manuscript.
This paper includes data collected by the \kepler\ mission. Funding for the \kepler\ mission is provided by the NASA Science Mission directorate. 
Work by K.M. was performed under contract with the California Institute of Technology (Caltech)/Jet Propulsion Laboratory (JPL) funded by NASA through the Sagan Fellowship Program executed by the NASA Exoplanet Science Institute. 
Work by H.K. is supported by a Grant-in-Aid from Japan Society for the Promotion of Science (JSPS), Nos. JP17K14246, JP18H01247, and JP18H04577.
This work was also supported by the JSPS Core-to-Core Program ``Planet$^2$" and the Einstein Fellowship Program executed by the Smithsonian Astrophysical Observatory.
The authors wish to recognize and acknowledge the very significant cultural role and reverence that the summit of Mauna Kea has always had within the indigenous Hawaiian community.  We are most fortunate to have the opportunity to conduct observations from this mountain.





\bibliographystyle{aasjournal}


\listofchanges

\end{document}